\documentclass[a4paper,fleqn]{cas-sc}

\usepackage[authoryear]{natbib}

\usepackage{ulem}

\def\tsc#1{\csdef{#1}{\textsc{\lowercase{#1}}\xspace}}
\tsc{WGM}
\tsc{QE}

\begin{document}
\let\WriteBookmarks\relax
\def\floatpagepagefraction{1}
\def\textpagefraction{.001}

\shorttitle{Phase lag enhances synchronization in coupled oscillators with inertia}

\shortauthors{Yi et al.}

\title [mode = title]{Phase lag enhances synchronization in coupled oscillators with inertia}

\author[1]{Sudo Yi}
\credit{Conceptualization, Investigation, Writing - original draft \& editing}

\author[1]{Cook Hyun Kim}
\credit{Conceptualization, Investigation, Writing - review \& editing}

\author[1]{Heetae Kim}
\cormark[1]
\ead{hkim@kentech.ac.kr}
\credit{Supervision, Funding acquisition, Writing - review \& editing}

\author[1]{B. Kahng}
\cormark[2]
\ead{bkahng@kentech.ac.kr}
\credit{Supervision, Project administration, Funding acquisition, Writing - review \& editing}

\affiliation[1]{organization={CCSS, KI for Grid Modernization, Korea Institute of Energy Technology},
            addressline={21 KENTECH-gil},
            city={Naju},
            postcode={58330},
            state={Jeonnam},
            country={Korea}}

\cortext[1]{Corresponding author}
\cortext[2]{Corresponding author}

\fntext[1]{}
\begin{abstract}
The second-order Kuramoto model with inertia exhibits different dynamical behaviors than the first-order KM without inertia. A central difference is its lower synchronization due to the emergence of multiple synchronized clusters with different frequencies. We aim to investigate how such lowered synchronization can be improved by applying external perturbations to the system in a steady state, for example, a symmetry-breaking phase lag to a subset of oscillators. We find that this phase lag steers the primary cluster along a specific path and enables it to merge with higher-order clusters, thereby enhancing global synchronization.
Our results reveal a mechanism by which controlled phase lag can improve entrainment in inertial oscillator systems, with possible implications for synchronization control in inertial oscillator networks.
\end{abstract}


\maketitle

\section{Introduction}\label{sec:introduction}
Synchronization is a fundamental phenomenon observed in a wide range of complex systems, including power grids, neural networks, and biological rhythms. To understand the collective behavior underlying such systems, the Kuramoto model (KM) has been widely employed as a theoretical framework for its simplicity and analytical tractability, offering valuable insights into the emergence of coherent phenomena such as macroscopic synchronization from microscopic interactions. 

In particular, stable operation of oscillating systems requires coherent behavior among numerous coupled rotating oscillators, such as generators and motors, even under continuous fluctuations in load and generation. To model such systems more realistically, the Kuramoto-like model with inertia, called the second-order Kuramoto model (2nd KM), has been widely adopted because it captures the essential mechanical dynamics of synchronous machines, in which inertia plays an important role in instantaneously recovering the system from fluctuations. The 2nd KM is expressed as 
\begin{equation}
m\ddot{\varphi}_i + D\dot{\varphi}_i = \omega_i +\frac{K}{N}\sum_j \sin(\varphi_j-\varphi_i),
\label{eq:2nd_KM}
\end{equation}
where $\varphi$, $\dot \varphi$, and $\ddot \varphi$ denote, respectively, phase, angular velocity, and angular acceleration. $m$, $D$, $K$, and $N$ denote, respectively, the inertia, damping coefficient, coupling strength, and total number of oscillators in the system. The quantity $\omega_i$ represents the intrinsic (natural) frequency of oscillator $i$, drawn from the normal distribution $g(\omega)$. For simplicity, we set $D=1$, with the remaining parameters understood to be rescaled accordingly with respect to $D$.
Unlike the conventional 1st KM without inertia, the 2nd KM hampers synchronization due to the kinetic energy of oscillators~\cite{Tanaka1997,Tanaka1997_2,Gao2018,Gao2021,Kim2025}.

The 2nd KM displays characteristics that are clearly distinct from those of the 1st KM: its synchronization transition is discontinuous, leading to the emergence of a hysteresis loop. Consequently, the transition point in the forward direction is delayed relative to that in the backward direction. In addition, the 2nd KM shows not only a primary synchronized cluster but also a couple of secondary and higher-order frequency-locked synchronized clusters. Each of them evolves with its own phase velocity~\cite{Gao2021,Kim2025}. These clusters make control and stabilization more challenging, creating substantial obstacles for controlling inertial oscillator systems, including simplified models inspired by power-grid synchronization.

To address this issue, recent works have introduced various modifications, including network topology optimization~\cite{Pinto2016}, higher-order coupling schemes~\cite{Jaros2023}, and time-delay-based coupling adjustments~\cite{Prousalis2022}. In this paper, we demonstrate that introducing a phase lag to a fraction of oscillators in the 2nd KM (with details below) offers an alternative strategy. The central finding of this work is that a phase lag, which is usually regarded as a source of frustration and hence as a desynchronizing perturbation, can instead enhance synchronization in the inertial Kuramoto model when the system already supports multiple frequency-locked clusters for large inertia. This enhancement does not arise from a uniform increase of coherence within a cluster, but from a dynamical reorganization of the cluster structure. The primary synchronized cluster is shifted asymmetrically and absorbs secondary locked clusters and nearby drifting oscillators without losing any of its originally locked oscillators. This enhancement is not possible in the first-order Kuramoto model due to the absence of multiple synchronized clusters with different frequencies.

\section{Model}\label{sec:Model}
We modify the 2nd KM by introducing a phase lag to a randomly selected $\eta$ fraction of oscillators after the system reaches a steady state without the phase lag. The equation of motion of modified 2nd KM is written as   
\begin{equation}
m\ddot{\varphi}_i + D\dot{\varphi}_i = \omega_i +\frac{K}{N}\sum_j \sin(\varphi_j-\varphi_i-\alpha_i),
\label{eq:2nd_KM_modified}
\end{equation}
where $\alpha_i\in [0,\pi/2)$ represents the phase lag. These phase lags are assigned to a randomly selected fraction $\eta$ of oscillators. It is formulated as 
\begin{equation}
    f(\alpha) = \eta \delta(\alpha - \alpha_0) + (1 - \eta)\delta(\alpha),
    \label{eq:falpha}
\end{equation}
where a fraction $\eta$ of oscillators is assigned the phase lag $\alpha_0$ in the coupling term, while the remaining fraction $1-\eta$ has no phase lag. This setup allows us to study how the phase lag influences synchronization and cluster formation. The stored data for each $K$ is used as the initial condition for the evolution of the modified KM. Despite the presence of the phase lag, the coupling remains attractive, which means that synchronization is still favored for sufficiently large $K > K_c$.

The use of a node-dependent phase lag is intended as a minimal model of controlled heterogeneity. In contrast to the standard Kuramoto--Sakaguchi model, where the same phase lag is imposed on all interactions, our setup represents a situation in which only a subset of oscillators is subject to a phase-lagged response. Such a perturbation may arise from heterogeneous local control, actuator-induced phase shifts, or phase-shifting devices in oscillator networks. We choose a binary distribution of $\alpha_i$ not because it is the most general case, but because it isolates the essential effect of symmetry breaking in the simplest possible form.

The synchronization process of the 2nd KM depends on initial configurations. It is especially sensitive when $K$ is increased. We numerically implement a synchronization transition by quasi-statically increasing $K$. In the beginning, a phase $\varphi_i$ of each oscillator $i$ is assigned a randomly selected value from the interval $[0,2\pi)$, and its angular velocity $\dot \varphi_i=\omega_i$. The system first evolves following Eq.~\eqref{eq:2nd_KM} without a phase lag from $K=0.01$ to higher $K$ values in steps of $\Delta K = 0.01$, storing each steady-state configuration.

\section{Results}\label{sec:Results}
\begin{figure}
  \centering
  \includegraphics[width=0.8\textwidth]{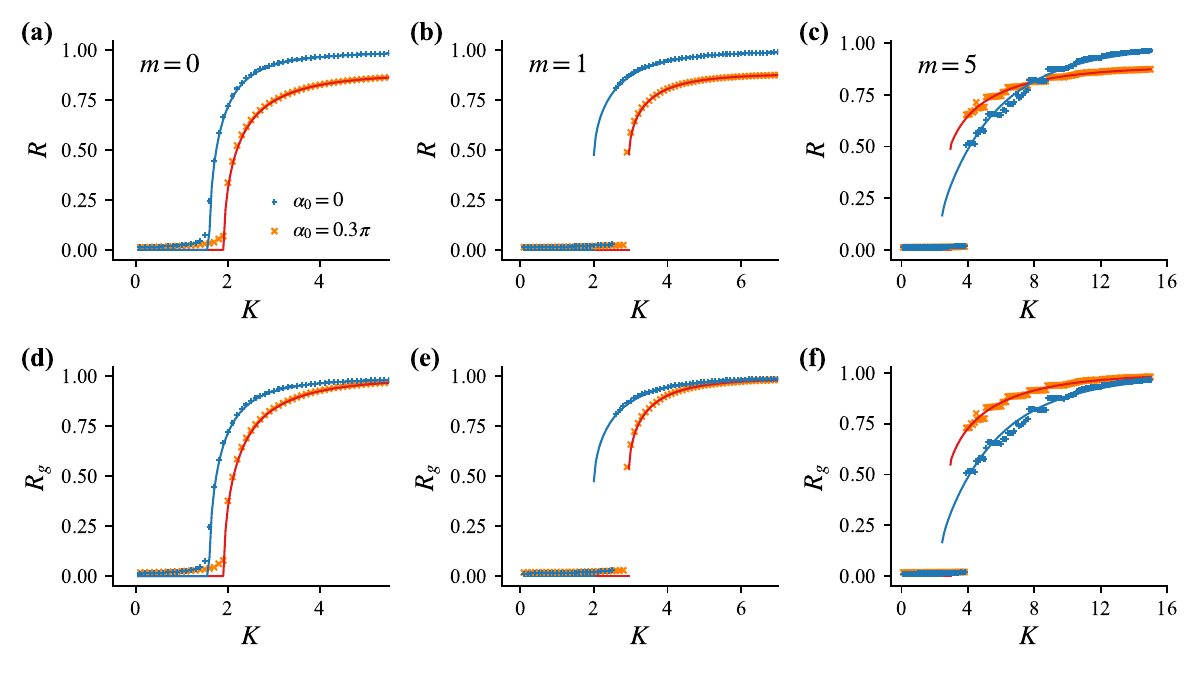}
    \caption{Plot of the order parameter $R$ for the 2nd KM without (blue) and with (orange) phase lags as a function of the coupling constant $K$ for (a) $m=0$, (b) $m=1$, and (c) $m=5$ for a single realization (left column).
For $m=0$ and $m=1$, imposing a phase lag reduces $R$, whereas for $m=5$, $R$ shows a crossover behavior as $K$ varies. Close to the transition point $K_c$, synchronization is enhanced, while for sufficiently large $K$ it is suppressed. Note also that the transition point $K_c$ is shifted to larger values when the phase lag is introduced. The enhanced synchronization occurs because, in case (c), the boundary of the primary cluster in the $\omega > 0$ region remains fixed [Fig.~\ref{fig:fig2}(e)], while the cluster itself extends into the $\omega < 0$ region.
To emphasize this effect, the group-based order parameter $R_g$ is shown in the right column for (d) $m=0$, (e) $m=1$, and (f) $m=5$. For $m=5$, the synchronization measured by $R_g$ is enhanced for all $K$.
The numerical data (symbols) are consistent with the solutions of the self-consistency equation (solid lines).}
\label{fig:fig1}
\end{figure}

\begin{figure}
  \centering
  \includegraphics[width=0.8\textwidth]{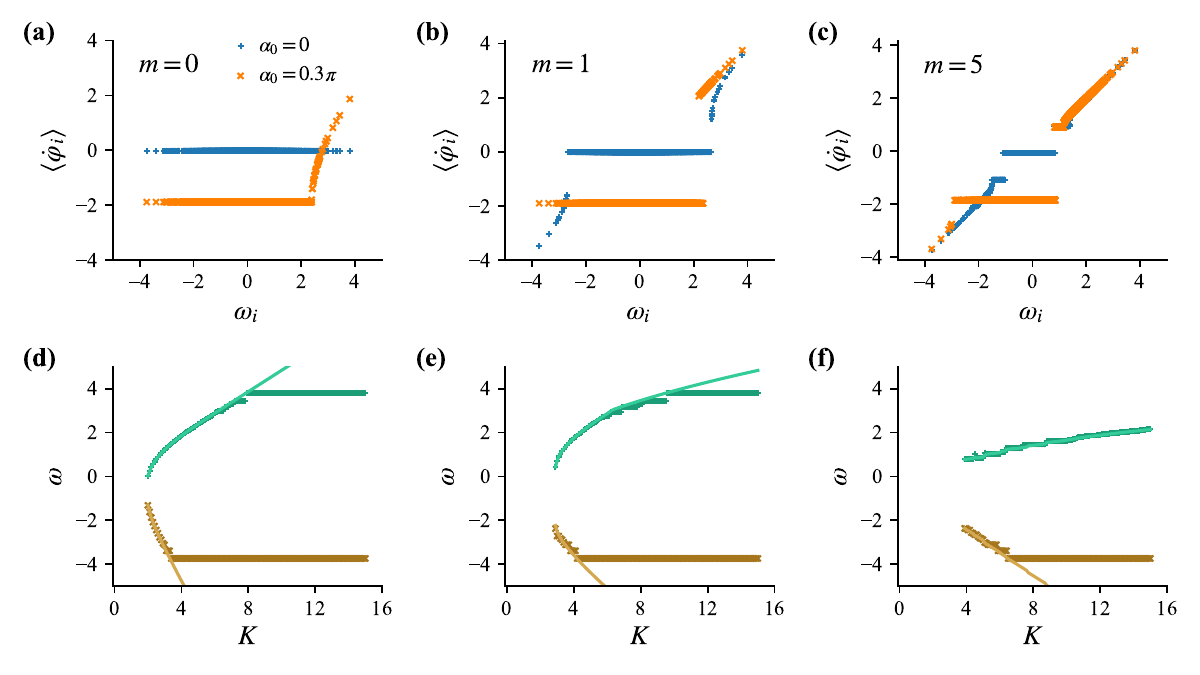}
    \caption{(Left column) Plots of the mean angular velocities of the primary and higher-order synchronized clusters versus intrinsic frequencies for (a) $m=0$, (b) $m=1$, and (c) $m=5$ with $N=5000$, and $K=5$, starting from the phase-lag-free state ($\alpha_0=0$) (blue) and after introducing the phase-lag with $\alpha_0=0.3\pi$ in the steady states (orange). In all cases, the phase lag effectively shifts the boundary of the primary cluster—defined by the region in which $\langle \dot{\varphi}_i\rangle$ is almost constant with respect to $\omega_i$—toward smaller values of $\omega_i$.
Yet, as illustrated for $m=5$ in (c), if the depinning threshold lies to the right of the original boundary, depinning does not occur and the right boundary is left unchanged.
To confirm this behavior, we plot the numerically obtained left and right boundaries (symbols), $\omega_l$ and $\omega_r$, as functions of $K$ for (d) $m=0$, (e) $m=1$, and (f) $m=5$, and compare them with the theoretical predictions (solid lines): $\omega_{l,r}=\Omega^R \mp KR$ for $m=0$ (see SM~\cite{SM} for details) and Eqs.~\eqref{eq:omega_l} and~\eqref{eq:omega_r} for $m>0$. In (d), (e), and (f), the apparent plateau of data points at large $\omega$ in the simulations is due to finite-size effects, because $g(\omega>|4|)\approx 6.33\times 10^{-5}<1/N$.}
\label{fig:fig2}
\end{figure}

We first examine how the phase lag affects the synchronization order parameter. The order parameter is enhanced only when the inertia is sufficiently large to generate multiple synchronized clusters. When $m=1$ and $\eta=0.5$, the inertia is too small for the system to form the secondary cluster, as shown in Fig.\ref{fig:fig1}(c). When $m=5$ and $\eta=0.5$, the strategy works, and the order parameter is enhanced near the transition point, as shown in Fig.\ref{fig:fig1}(e). Note that the transition point is slightly increased when the phase lag is imposed. The different results between the two cases with $m=1$ and $m=5$ originate from whether the secondary synchronized cluster forms or not, as shown in Fig.~\ref{fig:fig2}. Recently, it was shown that $m^*\approx 3.87$ is the minimum characteristic value for the formation of secondary synchronized clusters~\cite{kim2025_inertia}.


\begin{figure}
  \centering
  \includegraphics[width=0.48\textwidth]{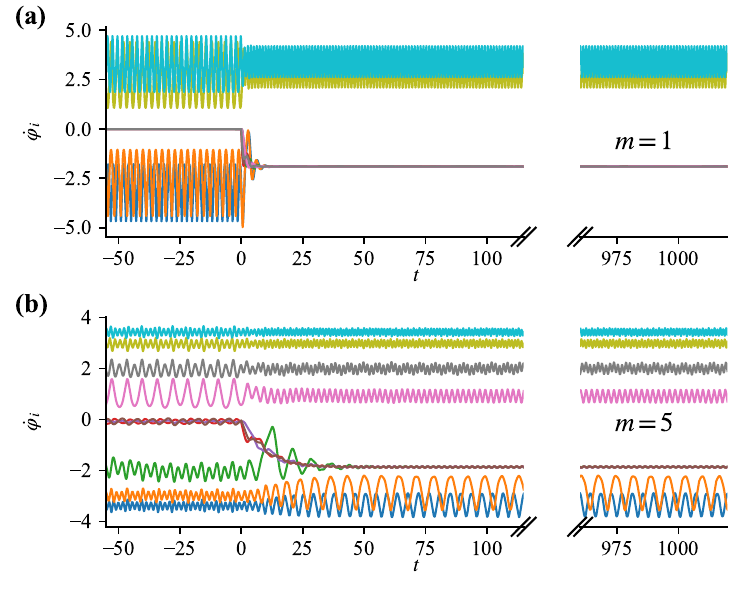}
    \caption{Time evolution of the instantaneous frequencies $\dot{\varphi}_i(t)$ of selected oscillators with different intrinsic frequencies $\omega_i$, (a) and (b) show the cases $m=1$ and $m=5$, respectively, with $N=5000$, $K=5$, $\alpha_0=0.3\pi$, and $\eta=0.5$. For $t<0$, the system evolves without phase lag and is in a steady state. At $t=0$, the phase lag is introduced to an $\eta$ fraction of oscillators. The system relaxes to a new steady state within a time scale much shorter than the transient time ($1000$) used in the simulations.}
\label{fig:varphi_versus_t}
\end{figure}

Figs.\ref{fig:fig2}(a), (b), and (c) display the mean angular velocity of each oscillator as a function of its intrinsic frequency for $K=5$ and several values of $m$. The curves are presented both before and after introducing the phase lags, with $\alpha_0=0.3\pi$ and $\eta=0.5$. The mean angular velocity is computed as $\langle \dot \varphi_i\rangle = (1/T)\sum_{t}^{T} \dot{\varphi}_i$ once a steady state is reached in the time window $t=[1000, 3000]$, so that $T=2000$ (see Fig.\ref{fig:varphi_versus_t}). Time counting starts when the phase lag is imposed. Therefore, the averaging interval lies entirely within the steady-state regime. This strategy differs from the previous approach~\cite{Tyloo2019}, in which the perturbation occurs in the form of brief strong pulses or slowly varying noise. 

When $m > m^*$, the angular velocity of the primary cluster changes to match that of the secondary cluster on one side with negative angular velocity, as shown in Fig.\ref{fig:fig2}(c). Then, the primary and secondary clusters are merged, yielding a larger primary cluster, as shown in Fig.\ref{fig:varphi_versus_t}(b). Note that the shift of the primary cluster is driven by the oscillators with phase lag. The rest of the oscillators without phase lag also shift due to strong coupling strength within the primary cluster. 

\section{Theory}\label{sec:Theory}
We determine the boundary of the updated primary cluster. By introducing the rotating-frame phase $\psi_i = \varphi_i-\Omega^R t$, where $\Omega^R$ is the mean rotation frequency, and defining the order parameter $R$ through the relation $Re^{{\rm i}\Psi} = \frac{1}{N} \sum_{i} e^{{\rm i}\psi_i}$, the equation of motion [Eq.~\eqref{eq:2nd_KM_modified}] reduces to
\begin{equation}    
m\ddot{\psi}_i + \dot{\psi}_i =\omega_i -\Omega^R +K R\sin(\Psi-\psi_i-\alpha_i).
\end{equation}
After the transformation $\theta_i=\psi_i-\Psi$ and the time rescaling $\tau=t\sqrt{KR/m}$, the equation becomes
\begin{equation}
\ddot{\theta}_i + a\dot{\theta}_i = b_i-\sin(\theta_i+\alpha_i),
\label{eq:reduced}
\end{equation} 
where $a={1}/{\sqrt{KRm}}$ and $b_i=(\omega_i-\Omega^R)/(KR)$. By the mapping $\theta_i\to\theta_i-\alpha_i$, Eq.~\eqref{eq:reduced} can be expressed in a simpler form as in the $\alpha_i=0$ frame, which is identical in form to the $\alpha_i=0$ case analyzed in Refs.~\cite{Tanaka1997,Tanaka1997_2}. Therefore, the Melnikov criterion that determines $b_s(a)$---the boundary between the coexistence region and the fixed-point-only region---applies without modification. The threshold is given approximately by
\begin{equation}
b_s(a) \simeq 
\begin{cases}
\frac{4}{\pi} a - 0.3056a^{3}, & 0 \le a \le a_{*}(1) \simeq 1.193, \\
1, & a \ge a_{*}(1),
\end{cases}\label{eq:EM_bsa}
\end{equation}
and the bistability window $b_s(a) < |b_i| < 1$ is independent of $\alpha_i$. For $|b_i| > 1$, only the limit cycle exists, while for $|b_i| < b_s(a)$, only the fixed point exists. 

Thus, the locked oscillators ($|b_i|<1$) satisfy the following equation
\begin{equation}
e^{{\rm i}\theta_i}e^{{\rm i}\alpha_i} = \sqrt{1-b_i^2}+{\rm i}b_i,
\label{eq:locked}
\end{equation}
and the drifting oscillators make the following approximation,
\begin{equation}
\langle e^{{\rm i}\theta_i}\rangle e^{{\rm i}\alpha_i} \approx \frac{1}{2}\left(-1+\frac{{\rm i}a^2}{b_i}\right)\frac{a^2}{a^4+b_i^2},
\label{eq:drifting}
\end{equation}
where this closed-form approximation is valid in the limits $a^2 \ll |b_i|$ or $a \gg 1$ as derived in~\cite{Gao2018}. Since the approximation is performed in the $\phi_i$ frame (where $\alpha_i$ is absent), the result is $\langle e^{{\rm i}\phi_i}\rangle$. Reverting to the original variable via $e^{{\rm i}\theta_i} = e^{-{\rm i}\alpha_i}\,e^{{\rm i}\phi_i}$ gives Eq.~\eqref{eq:drifting}. 

The introduction of phase lag shifts the cluster boundaries asymmetrically. On the left side ($\omega < \Omega^R$), the oscillators in limit cycles are captured by the primary cluster; this process is referred to as the pinning process. An oscillator on a limit cycle is captured by the fixed point when $|b_i|$ decreases below $b_s(a)$. Using $b_i = (\omega_i - \Omega^R)/(KR)$ and the leading-order expansion of $b_s(a)$ [Eq.\eqref{eq:EM_bsa}], the pinning condition $b_i = -b_s(a)$ (for the $\omega_i < \Omega^R$ side) translates directly to 
\begin{equation}
    \omega_l \equiv \Omega^R - \frac{4}{\pi}\sqrt{\frac{KR}{m}} + \frac{0.3056}{m}\frac{1}{\sqrt{KRm}}.
    \label{eq:omega_l}
\end{equation}
Because $b_s(a)$ is independent of $\alpha_i$, $\omega_l$ depends on the phase lag only through the global quantities $\Omega^R$ and $R$, which themselves change when the lag is introduced. The resulting boundary is expressed below~\cite{Gao2021}. 

On the right side, oscillators locked to the fixed point escape to a limit cycle when $|b_i|$ exceeds unity when the phase lag is applied. The depinning condition $b_i > 1$ gives
\begin{equation}
\omega_r^{(\mathrm{depinning})} = \Omega^R + KR.
\end{equation}
However, when the system is initialized from a lag-free steady state, the right boundary cannot exceed the original lag-free boundary
\begin{equation}
\omega_r^{(\mathrm{lag\text{-}free})} = \Omega_0^R + \frac{4}{\pi}\sqrt{\frac{KR_0}{m}} - \frac{0.3056}{m}\frac{1}{\sqrt{KR_0m}},
\end{equation}
with $R_0$ and $\Omega_0^R$ being the order parameter and mean frequency in the absence of phase lag because no oscillators beyond this point were locked in the first place. Hence, the effective right boundary is written as
\begin{equation}
\omega_r \equiv \text{min}\!\left(\omega_r^{(\mathrm{depinning})},\;
\omega_r^{(\mathrm{lag\text{-}free})}\right).
\label{eq:omega_r}
\end{equation}
For large inertia $m$, the lag-free cluster is narrow enough that $\omega_r^{(\mathrm{depinning})}$ typically exceeds $\omega_r^{(\mathrm{lag\text{-}free})}$, so no depinning occurs and the right boundary remains fixed [Fig.~\ref{fig:fig2}(f)].

The role of the phase lag can be understood as an effective bias acting on the force balance of each oscillator. In the rotating frame, the phase lag shifts the stable fixed point of the tilted washboard potential without changing the Melnikov threshold itself. Therefore, the local locking condition remains governed by the same bistability structure as in the lag-free system, whereas the global quantities \(R\) and \(\Omega^R\) are reorganized. This produces an asymmetric movement of the primary cluster boundaries: pinning on the left and possible depinning on the right. In the large-inertia regime studied here, however, depinning is suppressed because the depinning threshold lies beyond the original lag-free boundary. The coexistence of these two processes is a hallmark of hysteretic synchronization under inertia. The boundaries of the primary cluster obtained from the above expressions show good agreement with the simulation results, as illustrated in Fig.~\ref{fig:fig2}.

By combining the two contributions in Eqs.~\eqref{eq:locked} and \eqref{eq:drifting} with the distributions $f(\alpha)$ and $g(\omega)$, we obtain the following self-consistency equation for $R$ (see SM for the derivation):
\begin{equation}
R = \int e^{-{\rm i}\alpha} f(\alpha) \left[ \int_{\omega \in (\omega_l, \omega_r)} (\sqrt{1-b^2} + {\rm i}b)\, g(\omega)\, d\omega
+ \int_{\omega \notin (\omega_l,\omega_r)} \frac{1}{2}\left(-1 + \frac{{\rm i}a^2}{b}\right) \frac{a^2}{a^4 + b^2} g(\omega) d\omega \right] d\alpha,
\label{eq:self}
\end{equation}
where \((\omega_l, \omega_r)\) specifies the interval of locked oscillators. Note that the order parameter $R$ receives contributions from not only locked oscillators [Eq.~\eqref{eq:locked}] in the interval $(\omega_l, \omega_r)$, but also drifting oscillators [Eq.~\eqref{eq:drifting}] outside this interval. The numerical values of $R$ obtained from the self-consistency Eq.~\eqref{eq:self} are plotted as solid curves in Fig.~\ref{fig:fig2}(d), (e), and (f). These results agree with the solutions from the modified 2nd KM integrated using the fourth-order Runge–Kutta method, except in the vicinity of the transition point, where finite-size effects become relevant~\cite{Gao2018,Olmi2014}.

Introducing a phase lag can lead to an enlargement of the primary cluster, and this effect becomes more significant for large values of $m$. This indicates that only a small number of oscillators depin when the phase lag is applied. Consequently, the right boundary of the cluster remains fixed, while the left boundary shifts further to the left, resulting in an overall expansion of the primary cluster.

Synchronization can be characterized by the Kuramoto order parameter $R$. Furthermore, 
if oscillators are grouped according to whether or not they are subject to a phase lag with Eq.~\ref{eq:falpha}, the order parameter of synchronization within each group—measured by $R_g$, defined as 
\begin{equation}
R_g e^{{\rm i}\Psi_g} = \frac{1}{N_g} \sum_{j \in g} e^{{\rm i}\varphi_j},
\label{eq:def_Rg}
\end{equation}
with $N_g=\eta N$ or $(1-\eta)N$, depending on whether the group is subject to a phase lag. Because Eq.~\eqref{eq:reduced} can be mapped to the same reduced equation for every oscillator by absorbing $\alpha_i$ into the phase variable, each group follows identical dynamics in its own shifted frame. Therefore, the group order parameter $R_g$ takes the same value in each group, independently of $\eta$.

This naturally leads to the question of how such behavior affects phase synchronization. We find that $R_g$ is enhanced compared to the case without a phase lag, as shown in Fig.~\ref{fig:fig1}(f) when $m > m^*$. This is due to the incorporation of higher-order clusters located on one side of the primary cluster. It is important to check whether the order parameter $R$ behaves similarly because it is obtained by combining two contributions whose phases differ by $\alpha_0$, namely $R = R_g \, \big| \eta + (1 - \eta)e^{{\rm i}\alpha_0} \big|$. Thus, even though $R_g$ is enhanced in the presence of a phase lag, the global synchronization $R$ can be reduced due to the phase difference between the two subgroups, which can be seen in Fig.\ref{fig:fig1}(c). 

\section{Conclusion}
In summary, we have shown that imposing a phase lag on a subset of oscillators can enhance synchronization. This occurs in a regime where the 2nd KM system already forms the primary and higher-order synchronized clusters. Concretely, we first let these clusters emerge in the original 2nd KM without a phase lag in a steady state. Afterward, we introduce the phase lag to a fraction of the oscillators and allow the system to evolve until it reaches another steady state. Then, we found that the order parameter increased. This synchronization-promoting effect is in stark contrast to the typically degrading influence of phase lag, as seen in the 1st KM with $m=0$ in Fig.\ref{fig:fig1}(a).

In the low-inertia regime ($m=1$), where only the primary cluster forms, the order parameter displays a discontinuous jump, while secondary clusters have not developed. This behavior is characteristic of explosive synchronization transitions~\cite{PhysRevLett.106.128701,PhysRevE.86.016102}. When a phase lag is introduced in this setting, the primary cluster grows in size and the transition is postponed; nevertheless, this effect is insufficient for the order parameter to compensate for the decrease caused by shifting the transition point.

In the large-inertia regime ($m=5$), the phase lag enlarges the primary cluster by drawing in additional oscillators, including those belonging to secondary and higher-order clusters, as well as oscillators in isolated limit-cycle states, all without causing any previously locked ones to detach. 
As a result, the group-based order parameter $R_g$ increases substantially, indicating that a partial phase lag can enhance coherence. The incorporation of secondary and higher-order clusters into the primary cluster is a key factor in improving synchronization. This is achieved by introducing a symmetry-breaking external perturbation in the form of a phase lag. Without these secondary and higher clusters, synchronization would not be enhanced, as shown in Fig.~\ref{fig:fig2}(b). To simplify, we consider the minimal case with uniform inertia $m > m^*$ and uniform damping (with $D=1$ after rescaling). This framework allows us to clearly identify the mechanism by which phase lag can induce cluster merging. Therefore, the present results should be viewed as a minimal mean-field demonstration of how phase lags can reorganize hysteretic cluster dynamics. 

Recent advances in synchronization further support this perspective, suggesting that enhancing damping, rather than inertia, can play a decisive role in achieving stable collective behavior~\cite{Sajadi2022}. This is because the boundary of the primary cluster is determined by the Melnikov criterion~\cite{melnikov1963stability,guckenheimer2013nonlinear}, which is governed by the balance between the energy non-conserving terms, namely damping and intrinsic frequency terms. In the present work, we further demonstrate that the introduction of phase lags indirectly reshapes this balance by modifying the global quantities $R$ and $\Omega^R$. As a result, the Melnikov-based boundary of the primary cluster shifts through pinning on one side and depinning on the other.

More broadly, our findings extend the theoretical understanding of the Kuramoto model with inertia by revealing how controlled heterogeneity modifies the balance between synchronization and desynchronization tendencies. Although the present study is formulated within a globally coupled mean-field framework, the mechanism identified here is not specific to a particular physical realization. Rather, it is associated with the emergence of multiple frequency-locked clusters and hysteretic pinning–depinning dynamics induced by interactions with phase lags. An important direction for future work is to determine whether, and under what conditions, this mechanism persists in sparse networks, systems with heterogeneous inertia or damping, or models with alternative phase lag distributions.

\section*{Acknowledgments}
This work is supported by the National Research Foundation of Korea with Grant No. RS-2022-NR071795 for HK, and with Grant No. RS-2023-00279802 for BK,  and KENTECH Research Grant No. KRG-2021-01-007 for BK.

\printcredits

\bibliographystyle{cas-model2-names}

\bibliography{ref}

\end{document}